\documentclass[twocolumn,aps,prl,superscriptaddress]{revtex4-1}%
\usepackage{amsmath}
\usepackage{amssymb}
\usepackage{amsfonts}
\usepackage{revsymb}
\usepackage{graphicx}
\usepackage{color}%
\usepackage{hyperref}

\providecommand{\U}[1]{\protect\rule{.1in}{.1in}}
\newtheorem{lemma}{Lemma}[section]
\newtheorem{theorem}[lemma]{Theorem}

\newtheorem{definition}[lemma]{Definition}

\begin{document}

\title{Device-independent randomness extraction for arbitrarily weak min-entropy source}
\author{Jan Bouda}
\affiliation{Faculty of Informatics, Masaryk University,  Botanick\'a 68a, 602 00 Brno, Czech Republic}
\affiliation{F\'isica Te\'orica: Informaci\'o i Fen\'omens Qu\'antics
 Universitat Aut\'onoma de Barcelona, 08193 Bellaterra (Barcelona), Spain}
\affiliation{LIQUID: Lepanto Institute for Quantum Information and
Decoherence, Carrer de Lepant 307, 08025 Barcelona, Spain}
\author{Marcin Pawlowski}
\affiliation{Instytut Fizyki Teoretycznej i Astrofizyki, Uniwersytet
Gda\'nski, PL-80-952 Gda\'nsk, Poland}
\affiliation{School of Mathematics, University of Bristol, Bristol BS8 1TW, United Kingdom}
\author{Matej Pivoluska}
\affiliation{Faculty of Informatics, Masaryk University,  Botanick\'a 68a, 602 00 Brno, Czech Republic}
\author{Martin Plesch}
\affiliation{Faculty of Informatics, Masaryk University,  Botanick\'a 68a, 602 00 Brno, Czech Republic}
\affiliation{Institute of Physics, Slovak Academy of Sciences,
Bratislava, Slovakia}

\begin{abstract}
Expansion and amplification of weak randomness plays a crucial role in many
security protocols. Using quantum devices, such procedure is possible even
without trusting the devices used, by utilizing correlations between outcomes
of parts of the devices. We show here how to extract random bits with an
arbitrarily low bias from a single arbitrarily weak min-entropy source in
a device independent setting. To do this we use Mermin devices that exhibit
super-classical correlations. Number of devices used scales polynomially in
the length of the random sequence $n$. Our protocol is robust, it can tolerate devices that malfunction with a probability dropping polynomially in $n$ at the cost of a minor increase of the number of devices used.

\end{abstract}
\maketitle

High quality randomness is a very useful resource in many computation and
cryptographic tasks. In fact it has been shown that many protocols (including
quantum ones) vitally require perfect randomness for their security\cite{McInnesPinkas-ImpossibilityofPrivate-1991, BoudaPivoluskaPleschEtAl-Weakrandomnessseriously-2012, HuberPawlowski-Weakrandomnessin-2013}.

Unfortunately, at the same time perfect randomness is very rare. In the classical
world the true randomness, i.e. independent uniformly distributed random bits,
cannot be produced at all. The only available resource is pseudo-randomness,
sequences that appear random to all observers (often referred to as adversaries) not having full information about
the whole environment. Thus classical randomness generators produce pseudorandom numbers stemming from external sources and fluctuations, hoping that
the adversary will not be able to reconstruct all the background information.
Sources producing imperfect randomness even taking
into account the limited capabilities of the adversary are called weak random
sources. 
To enhance the quality and security of these sources, randomness
extractors are used. These are devices that combine more sources of randomness
to obtain fewer bits of higher quality \cite{Shaltiel-IntroductiontoRandomness-2011}.

On the other hand, theoretically the production of true randomness is possible, if one assumes Quantum theory to be valid: Preparation of a pure
state and measurement in its complementary basis will yield a perfectly random
result. This is due to the inherent randomness present in Quantum theory
itself - this principle is being used in the design commercially available devices \cite{-IDQuantique:-}.  The assumption, however, is high quality and stability of quantum devices in an adversarial setting, which is far from trivial to achieve \cite{Solc`a-Testingofquantum-2010}.

In addition, quantum devices in reality act more like black boxes that are
inaccessible for users except for providing them classical inputs and
obtaining classical outputs from them. It is very hard, if not impossible, to directly
test what these devices actually do, whether they perform operations and
measurements as promised and whether their outputs really come from quantum
measurements. Therefor it is crucial to test these devices
even during their activity - satisfying these tests shall guarantee that the
devices are correctly designed and manufactured and they work as
desired. This is possible by utilizing super-classical correlations of certain
quantum states - if the device consists from separate parts, their classical results
can be tested for correlations and their level, if breaking the classical
bound, can be a guarantee of their quantum nature. Using non-trusted (or
self-testing) quantum devices is referred as Device independence in a broader
scope. The process of transformation of a weak random source into uniformly
random bits is called \emph{randomness extraction} throughout this letter.

\textit{Weak random sources --} To provide a figure of merit of randomness
extractors, one needs to characterize the randomness of the input random
source. One of the possible parameterizations is the so called
Santha--Vazirani (SV) parametrization
\cite{SanthaVazirani-Generatingquasi-randomsequences-1986}, given by the
following property: Let $X=(X_{1},X_{2},\dots)$ be an arbitrarily long random
bit string produced by an $\varepsilon$-SV source. Then for any $1\leq i\leq
n$ it holds that
\begin{align}
&  \forall x_{1},\dots,x_{i-1}\in\{0,1\},\forall e\in{\mathcal{I}}(E),\\
&  \left\vert P(X_{i}=0|X_{i-1}=x_{i-1},\dots,X_{1}=x_{1},E=e)-\frac{1}%
{2}\right\vert \leq\varepsilon,\nonumber
\end{align}
where $E$ is any information an adversary Eve might hold. Note here that the
apparent randomness (i.e. without knowledge of $E$) of each $X_{i}$ may as
well be uniform. The purpose of introducing random variable $E$ is to
represent possible correlations between the choice of the measurement settings
and internal workings of the devices running a Bell type test.

Second possibility is to consider a one-shot use source that would produce
$n$-bit strings $X$ (with $n$ being arbitrary large). Here we can characterize
the randomness of the source by the (conditional) min-entropy of the produced sequence
defined as
\[
H_{\infty}(X\vert E)=-\log_{2}\max_{x\in\mathcal{I}(X),e\in\mathcal{I}(E)}P(X=x\vert E=e).
\]
A source is called an $\left(  n,k\right)  $ source if $H_{\infty}(X\vert E)\geq k$
and might be also characterized by its min-entropy rate $R=k/n$.

Combining these two approaches we get the reusable min-entropy source with $n$-bit blocks of
output with guaranteed min-entropy $k$. Such a source can be modeled as a
sequence of $n$-bit random variables $X_{1},X_{2},\dots$, such that
\begin{align}
&  \forall x_{1},\dots,x_{i-1}\in\{0,1\}^{n},\forall e\in{\mathcal{I}%
}(E),\label{eq:min-entropy_definition}\\
&  H_{\infty}(X_{i}|X_{i-1}=x_{i-1},\dots,X_{1}=x_{1},E=e)\geq k.\nonumber
\end{align}
Therefore, each new block has a guaranteed minimal min-entropy, even conditioned on the
previous ones and any information of the adversary. It is easy to see that SV
sources are recovered with $n=1$ and $\varepsilon=2^{-H_{\infty}(X)}-\frac
{1}{2}$. Source of this type is also called \emph{block source}.

Classically the task of transforming a single weak source, characterized
either as a Santha-Vazirani source, or a min-entropy (block) source into a
fully random bit is known to be impossible
\cite{SanthaVazirani-Generatingquasi-randomsequences-1986,Shaltiel-IntroductiontoRandomness-2011}%
. However, with non-classical resources the task becomes possible. More
precisely, weak random source can be used to choose measurements for a Bell
test in order to certify that observed correlations cannot be explained by
local theories and thus must necessarily contain intrinsic randomness.

In their seminal paper Colbeck and Renner
\cite{ColbeckRenner-Freerandomnesscan-2012} showed that amplification of
Santha-Vazirani sources is possible for a certain range of parameter
$\varepsilon$ and thus opened a line of research devoted to SV amplification.
Subsequent works provided protocols that are able to amplify SV-sources for
any $\varepsilon<\frac{1}{2}$ in various settings
\cite{GallegoMasanesEtAl-Fullrandomnessfrom-2013,MironowiczPawlowski-Amplificationofarbitrarily-2013,
GrudkaHorodeckiHorodeckiEtAl-Freerandomnessamplification-2013,RamanathanBrandaoGrudkaEtAl-RobustDeviceIndependent-2013}%
. This line of researched culminated in the work of Brand\~{a}o~et.~al.
\cite{BrandaoRamanathanGrudkaEtAl-RobustDevice-IndependentRandomness-2013},
who showed how to amplify such source of randomness with the use of only eight
non-communicating devices. Their work was quickly followed by that of Coudron
and Yuan \cite{CoudronYuen-InfiniteRandomnessExpansion-2013}, who showed how
to use $20$ non-communicating devices to obtain arbitrary many bits from a
Santha-Vazirani source.

On the other hand, extraction from min--entropy sources is relatively
unexplored. There is a sequence of works exploring the validity of Bell tests
if the measurements are chosen according to a min--entropy source
\cite{KohHallSetiawanEtAl-EffectsofReduced-2012,ThinhSheridanScarani-Belltestswith-2013}
and the authors of this paper provided a protocol which uses $3$-party
GHZ-paradox to amplify sources with min-entropy rate $R>\frac{1}{4}\log
_{2}(10)$ against quantum adversaries
\cite{PleschPivoluska-SingleMin-EntropyRandom-2013}. Recently an extensive
work on this topic was made public on pre-print archive \cite{ChungShiWu-PhysicalRandomnessExtractors-2014}. In this
letter we conclude this work by providing a protocol extracting random bits
from min-entropy sources of randomness with any non-zero min--entropy rate.


\textit{Device-independent concept and Mermin inequality --}
In this letter we use the three partite Mermin inequality. Let's consider
three spatially-separated boxes, each of them having a single bit input and a
single bit output. Let us denote the input bits of the respective boxes by
$X,$ $Y$ and $Z$ and the corresponding output bits $A$, $B$ and $C$. By
construction we guarantee $X\oplus Y\oplus Z=1$, i.e. we consider only inputs
$XYZ\in\{111,100,010,001\}$ simultaneously passed to all boxes. The value $v$
of the Mermin term is a function of the $4$ conditional probabilities defined
by the behaviour of the device and of the probability distribution $p$ on
inputs
\begin{align}
v=  &  P(A\oplus B\oplus C=1|XYZ=111)P(XYZ=111)+\nonumber\\
+  &  P(A\oplus B\oplus C=0|XYZ=100)P(XYZ=100)+\nonumber\\
+  &  P(A\oplus B\oplus C=0|XYZ=010)P(XYZ=010)+\nonumber\\
+  &  P(A\oplus B\oplus C=0|XYZ=001)P(XYZ=001). \label{eq:mermin}%
\end{align}
In particular, for the uniform input distribution we set
$P(XYZ=111)=P(XYZ=010)=P(XYZ=001)=P(XYZ=100)=\frac{1}{4}$ and denote the Mermin term by
$v_{u}$.

Assuming the uniform distribution on all four inputs, the maximal value of
$v_{u}$ achievable by a classical device \cite{Mermin1990} is $\frac{3}{4}$
(thus the Mermin inequality reads $v_{u}\leq\frac{3}{4}$) and there exists a
classical device that can make any $3$ conditional probabilities
simultaneously equal to $1$. In the quantum world we can achieve $v_{u}=1$ and
satisfy perfectly all $4$ conditional probabilities using the tripartite GHZ
state $\frac{1}{\sqrt{2}}({|000\rangle}+{|111\rangle})$ and measuring
$\sigma_{X}$ ($\sigma_{Y}$) when receiving $0$ ($1$) on input.

The beautiful property of the Mermin inequality is that the violation $v$
gives us directly the probability that the device passes a specific test
$A\oplus B\oplus C=X\cdot Y\cdot Z$. The probability of failing the test reads
$w=1-v$.

Mironowicz, Gallego and Pawlowski (MGP)
\cite{MironowiczPawlowski-Amplificationofarbitrarily-2013} showed the
following result:
Take a linearly ordered sequence of $k$ Mermin devices $D_{1}...D_{k}$ ($k$
being arbitrary) that have uniform distribution on inputs, and each device
knows inputs and outputs of its predecessors (for optional cheating purposes),
but devices cannot signal to its predecessors. Let us assume that the inputs
of devices are described by random variables $XYZ_{1},\dots,XYZ_{k}$, and the
outputs by $ABC_{1},\dots,ABC_{k}$. Then there exists a function
$f(\varepsilon)$ such that if the value of the Mermin variable
(\ref{eq:mermin}) using uniform inputs is at least $v_{u}\geq f(\varepsilon)$, then the output bit
$A_{k}$ has a bias at most $\varepsilon$ conditioned on the input and output
of all its predecessors and the adversarial knowledge. This function can be
lower bounded by a Semi-Definite Program (SDP) using any level of the
hierarchy introduced in \cite{Navascues2008}. By using the second level of the
hierarchy one can obtain the bound on $f(\varepsilon)$ as a function of
$\varepsilon$ shown in Fig.1. \begin{figure}[th]
\center
\resizebox{9cm}{!}{
\includegraphics{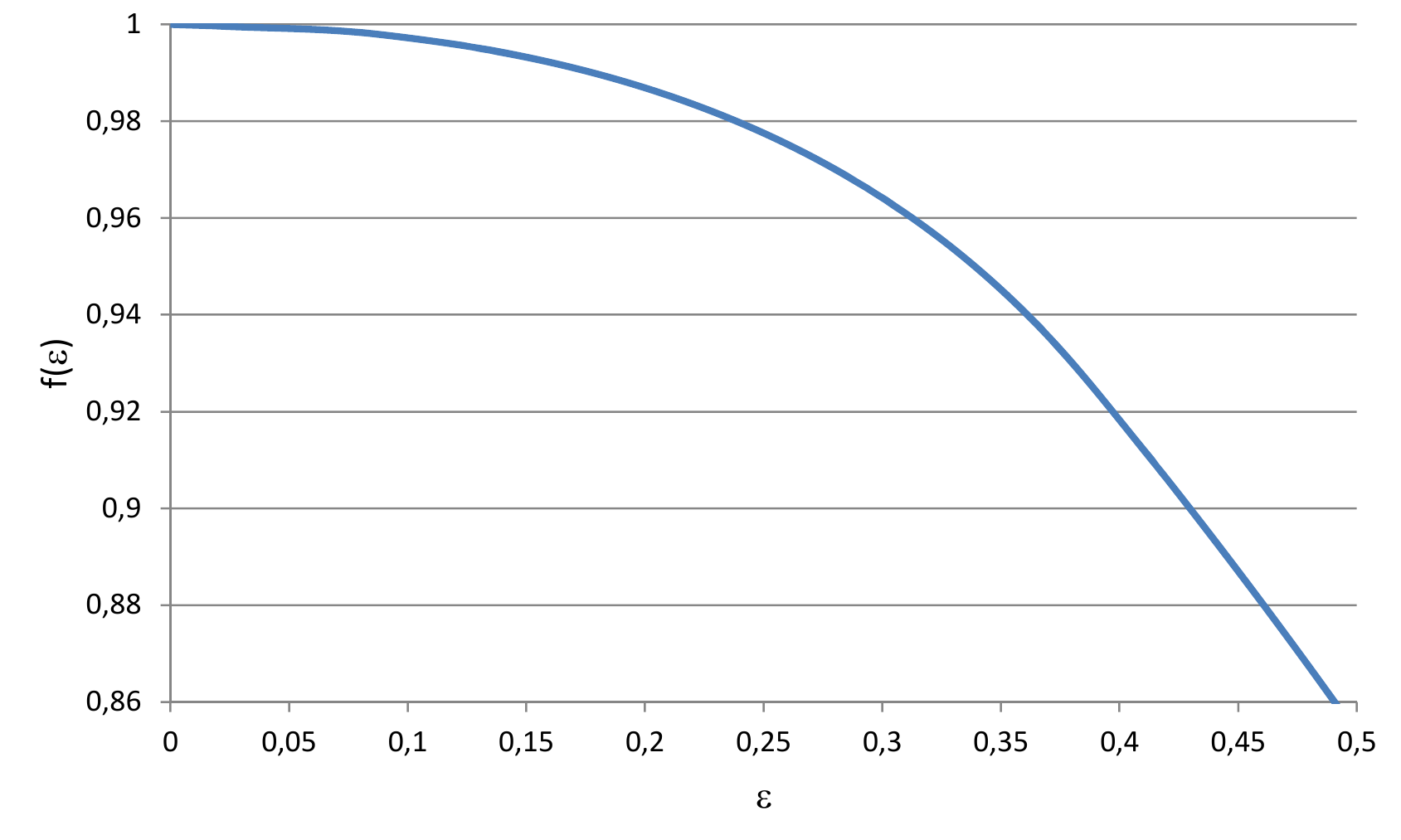}}\caption{Depicted is the value of Mermin
variable $v=f(\varepsilon)$ needed to certify the bias of the output bit to
be at most $\varepsilon$.}%
\end{figure}We can set $k=1$ (having just a single device) and get the lower
bound on the detection probability of producing a bit biased by more than
$\varepsilon$, which is $w_{u}>1-f(\varepsilon)$. More independent
non-communicating devices can be ordered into any sequence and thus this limit
holds for any of these devices simultaneously.


\textit{Single-round protocol --}
In the rest of our analysis we will be working with $(n,k)$ sources for an arbitrary $n$ and $k\geq 2$. This is to simplify the explanation, since by taking $\lceil\frac{2}{k'}\rceil$ blocks of an arbitrary $(n',k')$ source with $k'>0$ we get a $(n,k)$ source with $n=\lceil\frac{2}{k'}\rceil n'$ and $k=\lceil\frac{2}{k'}\rceil k'\geq 2$.

Let us start with a min-entropy $(n,2)$ source (recall that $(n,k)$ source
with $k>2$ is also an $(n,2)$ source) and define $N=2^{n}$. Let $H=\{h_{1}%
,\dots,h_{m}\}$ be a family of hash functions s.t. $h_{i}%
:\{0,...,N-1\}\rightarrow\{0,1,2,3\}$. Each hash-function $h_{i}$ is used to provide input for
a Mermin-type device $D_{i}$, where outputs of the function $0,1,2,3$
identify $111,100,010,001$ inputs for the device.

We want to construct $H$ with the property that for every $4$-element set
$S\subseteq\{0,\dots,N-1\}$ there exist at least one hash function $h\in H$
such that $h(S)=\{0,1,2,3\}$. This is trivially satisfied for the set of all
possible hashing functions $H_{full}=\{0,1,2,3\}^{N}$, however, such a class
of functions with its $4^{N}$ elements is unpractically large. In the
supplementary material we show a construction with logarithmic number of
functions in $N$, thus the number of devices needed scales polynomially with
the length of the sequence $n$. We also stress that for large $n$ one hash
function covers as many as $9\%$ of all four-tuples, independently on $n$. So
the size of an optimal set of hash functions might not depend on $n$ at all.

The protocol works as follows:

\begin{enumerate}
\item We obtain the (weakly) random $n$ bit string $X$ from the random number generator.

\item Into each device $D_{i}$ we input the $3$ bit string $h_{i}(X)$ --
inputs $X_{i}$, $Y_{i}$ and $Z_{i}$ and obtain the outputs $A_{i}$, $B_{i}$
and $C_{i}$.

\item We verify whether for each device $D_{i}$ the condition $Z_{i}\oplus
Y_{i}\oplus Z_{i}=A_{i}\cdot B_{i}\cdot C_{i}$ holds. If this is not true, we
abort the protocol due to cheating attempt of the provider.

\item We define the output bit of the protocol as $b=\bigoplus_{i=1}^{m}%
A_{i}.$
\end{enumerate}

The protocol is depicted in the Fig. 2.

\begin{figure}[th]
\begin{center}
\includegraphics[scale = 1.9]{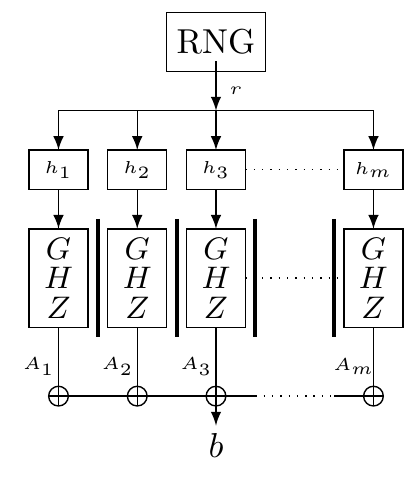}
\end{center}
\par
\label{figure:1round}\caption{Depiction of a single round protocol. Bit string
drawn from the flat random source is hashed into $m$ inputs into Mermin
devices so that at least one device receives perfectly random distribution.
This guarantees at least one result almost perfectly random, what also holds
for the product of individual results. }%
\end{figure}

Let us now examine the properties of the bit $b$. First consider only flat
$(n,2)$ distributions. Recall that these are exactly distributions that are
uniform on $4$-element subsets of the sample space. Our construction of the
class $H$ of hash functions assures that for any flat probability distribution
there is a function $h_{j}\in H$ and the corresponding device $D_{j}$ such
that inputs of $D_{j}$ (hashed by $h_{j}$) are uniform on this flat
distribution. This gives us that if adversary restricts himself to flat
distributions and wants to achieve bias greater than $\varepsilon$ for the output bit $b$, she must achieve this bias in all rounds.
The probability that she is not detected while doing this is $v_{u}\leq
f(\varepsilon)$ for each round. The same condition holds then also for the product of all
output bits $b$.

The set of all $(n,2)$ distributions is convex and flat distributions are
exactly all extremal points of this convex set. Thus any $(n,2)$ distribution
$d$ can be expressed as a convex combination of at most $N$ $(n,2)$ flat
distributions $d_{i}$ (Caratheodory theorem) as $d=\sum_{i=1}^{N}p_{i}d_{i}$
for some $p_{i}\geq0$, $\sum_{i=1}^{N}p_{i}=1$. The probability that the
adversary is not detected is given by the successful cheating probabilities
when using flat distribution $d_{i}\in\{d_{i}\}_{i=1}^{N}$ averaged thourgh
the probability distribution on these flat distributions $v_{u}\leq\sum
_{i=1}^{N}p_{i}P($not detected$|d_{i})\leq f(\varepsilon)\sum_{i=1}^{N}p_{i}$
\cite{ScaraniGisinBrunnerEtAl-Secrecyextractionfrom-2006}. Thus the upper bound $v_{u}\leq
f(\varepsilon)$ holds for non-flat distributions as well.

To summarize this part, having an $(n,k)$ source with $k\geq2$, with a single
round of a protocol, we can produce a single bit that is biased at most by
$\varepsilon$ with a certainty of
$1-f(\varepsilon)$.

\textit{Multiple-round protocol for block sources --} Let us state the most
general task: we have an $(n,k)$ block source with arbitrary $n$ and $k\geq 2$ (recall that any source with $k>0$ can be multiplied to obtain $k\geq 2$). We
would like to produce a bit that is biased by no more than $\varepsilon$ with
certainty of at least $1-\delta$.

If the one-round version does not meet these parameters, we will repeat the whole protocol $l$ times. By using
new devices and new outputs of the block source, each of the runs $j$ will
produce a bit $b_{j}$, that is biased by $\varepsilon$ from perfectly random
bit conditioned on all the previous bits up to a probability $f(\varepsilon)$.
Thus also the XOR of all output bits $b=%
{
\bigoplus\limits_{j=1}^{l}}
b_{j}$ will have at most the bias $\varepsilon$. After $l$ rounds, the
probability of the adversary not being detected will be upper bounded by
$f(\varepsilon)^{l}$. Note that the product form does not come from the fact that the detection probabilities are independent (they are not). This is a product of a chain of conditional probabilities. Recall that the bound $f(\varepsilon)$ holds conditioned by any inputs and outputs of the previous devices (in an arbitrarily ordering that respects the causality). Thus choosing $l>\frac{\log\delta}{\log f(\varepsilon)}$
will guarantee the fulfillment of the conditions for the parameters $\varepsilon$ and $\delta$.

Summing up, with an $(n,k)$ block source and $O\left(  \frac{\log\delta}{\log
f(\varepsilon)}Poly\left[  n\left\lceil \frac{2}{k}\right\rceil \right]
\right)  $ Mermin devices we can produce a single random bit with bias smaller
than $\varepsilon$ with probability largen than $1-\delta$. For producing more
bits we simply repeat the whole procedure: all the bits produced will have bias
smaller than $\varepsilon$ conditioned on the bits produced so far, with
linearly scaling of resources.

\textit{Protocol for one-shot min-entropy sources -- }We can
model a different scenario where the random source is described by a single use
min-entropy source characterized by its min-entropy rate $R$. In such a case
we cannot use the same scenario as before, as there are no independent blocks
of randomness with guaranteed min-entropy available. In spite of this fact
randomness extraction is still possible on the cost of increasing the number
of devices used.

We can draw a bit string from the source with length $n$ and min-entropy $Rn$,
securing at least $2^{Rn}$ realizations of the string appearing with non-zero
probability. We shall use this string for a single round of the protocol,
however using a full set of hashing functions $H_{full}$. Then, for flat sources, there will be
at least $\frac{Rn}{2}$ devices
obtaining perfectly random distribution on inputs independently on each other (see supplementary material for explicit construction), yielding failure
probability of the protocol $\delta<f(\varepsilon)^{\frac{Rn}{2}}$.
Thus choosing $n>\frac{2}{R}
\frac{\log(\delta)  }{\log(  f(  \varepsilon))}$ will produce a random bit biased by no more than $\varepsilon$
up to a probability $\delta$, though on the costs of double-exponential number
of devices in $\frac{1}{R}$ and $\frac{\log\left(  \delta\right)  }{\log f\left(  \varepsilon\right)  }$. For non-flat sources the same result holds due to Caratheodory theorem mentioned earlier.

\textit{Robustness --}
Aborting the protocol after even a single mistake of the devices is certainly
highly impractical from the imlementation point of view. Therefor we expand our
analysis into a situation where we tolerate certain noise on the devices,
which would manifest itself by occasional failing of the test condition even
for honest devices. More specifically, we shall tolerate a certain fraction of
the devices to malfunction without aborting the protocol.

In the supplementary material we show that we can tolerate up to
$l\frac{\left(  1-f(\varepsilon)\right)  }{2}\,\ $devices to fail in the whole
protocol and still achieve the same result as for the perfect protocol by
choosing $l>\frac{8\ln\delta}{f(\varepsilon)-1}.$ This translates into
increasing the number of rounds of the protocol comparing to the case of ideal
devices by a factor of $\frac{8\ln\left(  f(\varepsilon)\right)
}{f(\varepsilon)-1}$. For small $\varepsilon$ the parameter $f(\varepsilon)$
approaches $1$ and the multiplication factor saturates by $8$. For honest
devices with individual failure probability bounded by $\frac{\left(
1-f(\varepsilon)\right)  }{4m}$, the probability of a false alarm decreases
exponentially with the number of protocol rounds $l$.

\textit{Conclusion --}
In this letter we have introduced a protocol that extracts weak randomness
obtained from a min-entropy source in the device independent setting. The
protocol works for arbitrarily weak both single-use and block min-entropy
sources, with a reasonable scaling of the number of devices in the latter
case. Our protocol is also robust, as it allows tolerating some fraction of
malfunctioning devices at the cost of a constant increase of the number of
devices used.

\textit{Acknowledgements --} Authors thank P. Horodecki, A. Winter, and S.
Massar for insightful and stimulating discussions and Piotr Mironowicz for supplying the raw data for Fig.1. JB, MP2 and MP3 acknowledge
the support of the Czech Science Foundation GA CR project P202/12/1142 and
support of the EU FP7 under grant agreement no 323970 (RAQUEL). MP3
acknowledges VEGA 2/0072/12. JB acknowledges support by the European Research
Council through Advanced Grant "IRQUAT". MP1 acknowledges FNP TEAM, NCN grant
2013/08/M/ST2/00626 and ERC QOLAPS.

\bibliographystyle{apsrev4-1}
\bibliography{DIMinEntropExtract}

\begin{thebibliography}{22}%
\makeatletter
\providecommand \@ifxundefined [1]{%
 \@ifx{#1\undefined}
}%
\providecommand \@ifnum [1]{%
 \ifnum #1\expandafter \@firstoftwo
 \else \expandafter \@secondoftwo
 \fi
}%
\providecommand \@ifx [1]{%
 \ifx #1\expandafter \@firstoftwo
 \else \expandafter \@secondoftwo
 \fi
}%
\providecommand \natexlab [1]{#1}%
\providecommand \enquote  [1]{``#1''}%
\providecommand \bibnamefont  [1]{#1}%
\providecommand \bibfnamefont [1]{#1}%
\providecommand \citenamefont [1]{#1}%
\providecommand \href@noop [0]{\@secondoftwo}%
\providecommand \href [0]{\begingroup \@sanitize@url \@href}%
\providecommand \@href[1]{\@@startlink{#1}\@@href}%
\providecommand \@@href[1]{\endgroup#1\@@endlink}%
\providecommand \@sanitize@url [0]{\catcode `\\12\catcode `\$12\catcode
  `\&12\catcode `\#12\catcode `\^12\catcode `\_12\catcode `\%12\relax}%
\providecommand \@@startlink[1]{}%
\providecommand \@@endlink[0]{}%
\providecommand \url  [0]{\begingroup\@sanitize@url \@url }%
\providecommand \@url [1]{\endgroup\@href {#1}{\urlprefix }}%
\providecommand \urlprefix  [0]{URL }%
\providecommand \Eprint [0]{\href }%
\providecommand \doibase [0]{http://dx.doi.org/}%
\providecommand \selectlanguage [0]{\@gobble}%
\providecommand \bibinfo  [0]{\@secondoftwo}%
\providecommand \bibfield  [0]{\@secondoftwo}%
\providecommand \translation [1]{[#1]}%
\providecommand \BibitemOpen [0]{}%
\providecommand \bibitemStop [0]{}%
\providecommand \bibitemNoStop [0]{.\EOS\space}%
\providecommand \EOS [0]{\spacefactor3000\relax}%
\providecommand \BibitemShut  [1]{\csname bibitem#1\endcsname}%
\let\auto@bib@innerbib\@empty
\bibitem [{\citenamefont {McInnes}\ and\ \citenamefont
  {Pinkas}(1991)}]{McInnesPinkas-ImpossibilityofPrivate-1991}%
  \BibitemOpen
  \bibfield  {author} {\bibinfo {author} {\bibfnamefont {J.~L.}\ \bibnamefont
  {McInnes}}\ and\ \bibinfo {author} {\bibfnamefont {B.}~\bibnamefont
  {Pinkas}},\ }in\ \href {\doibase 10.1007/3-540-38424-3_31} {\emph {\bibinfo
  {booktitle} {Crypto'90}}}\ (\bibinfo {year} {1991})\ pp.\ \bibinfo {pages}
  {421--435},\ \bibinfo {note} {lNCS 537}\BibitemShut {NoStop}%
\bibitem [{\citenamefont {Bouda}\ \emph {et~al.}(2012)\citenamefont {Bouda},
  \citenamefont {Pivoluska}, \citenamefont {Plesch},\ and\ \citenamefont
  {Wilmott}}]{BoudaPivoluskaPleschEtAl-Weakrandomnessseriously-2012}%
  \BibitemOpen
  \bibfield  {author} {\bibinfo {author} {\bibfnamefont {J.}~\bibnamefont
  {Bouda}}, \bibinfo {author} {\bibfnamefont {M.}~\bibnamefont {Pivoluska}},
  \bibinfo {author} {\bibfnamefont {M.}~\bibnamefont {Plesch}}, \ and\ \bibinfo
  {author} {\bibfnamefont {C.}~\bibnamefont {Wilmott}},\ }\href {\doibase
  10.1103/PhysRevA.86.062308} {\bibfield  {journal} {\bibinfo  {journal} {Phys.
  Rev. A}\ }\textbf {\bibinfo {volume} {86}},\ \bibinfo {pages} {062308}
  (\bibinfo {year} {2012})}\BibitemShut {NoStop}%
\bibitem [{\citenamefont {Huber}\ and\ \citenamefont
  {Paw\l{}owski}(2013)}]{HuberPawlowski-Weakrandomnessin-2013}%
  \BibitemOpen
  \bibfield  {author} {\bibinfo {author} {\bibfnamefont {M.}~\bibnamefont
  {Huber}}\ and\ \bibinfo {author} {\bibfnamefont {M.}~\bibnamefont
  {Paw\l{}owski}},\ }\href {\doibase 10.1103/PhysRevA.88.032309} {\bibfield
  {journal} {\bibinfo  {journal} {Phys. Rev. A}\ }\textbf {\bibinfo {volume}
  {88}},\ \bibinfo {pages} {032309} (\bibinfo {year} {2013})}\BibitemShut
  {NoStop}%
\bibitem [{\citenamefont
  {Shaltiel}(2011)}]{Shaltiel-IntroductiontoRandomness-2011}%
  \BibitemOpen
  \bibfield  {author} {\bibinfo {author} {\bibfnamefont {R.}~\bibnamefont
  {Shaltiel}},\ }in\ \href {\doibase 10.1007/978-3-642-22012-8_2} {\emph
  {\bibinfo {booktitle} {Automata, Languages and Programming}}},\ \bibinfo
  {series} {Lecture Notes in Computer Science}, Vol.\ \bibinfo {volume}
  {6756},\ \bibinfo {editor} {edited by\ \bibinfo {editor} {\bibfnamefont
  {L.}~\bibnamefont {Aceto}}, \bibinfo {editor} {\bibfnamefont
  {M.}~\bibnamefont {Henzinger}}, \ and\ \bibinfo {editor} {\bibfnamefont
  {J.}~\bibnamefont {Sgall}}}\ (\bibinfo  {publisher} {Springer Berlin
  Heidelberg},\ \bibinfo {year} {2011})\ pp.\ \bibinfo {pages}
  {21--41}\BibitemShut {NoStop}%
\bibitem [{-ID()}]{-IDQuantique:-}%
  \BibitemOpen
  \href@noop {} {\enquote {\bibinfo {title} {Id quantique:},}\ }\bibinfo
  {howpublished}
  {\href{http://www.idquantique.com/random-number-generators/products/products%
-overview.html} {Quantis}}\BibitemShut {NoStop}%
\bibitem [{\citenamefont {Solc{\`a}}(2010)}]{Solc`a-Testingofquantum-2010}%
  \BibitemOpen
  \bibfield  {author} {\bibinfo {author} {\bibfnamefont {R.}~\bibnamefont
  {Solc{\`a}}},\ }\emph {\bibinfo {title} {Testing of a quantum random number
  generator}},\ \href@noop {} {Master's thesis},\ \bibinfo  {school} {ETH
  Z{\"u}rich} (\bibinfo {year} {2010})\BibitemShut {NoStop}%
\bibitem [{\citenamefont {Santha}\ and\ \citenamefont
  {Vazirani}(1986)}]{SanthaVazirani-Generatingquasi-randomsequences-1986}%
  \BibitemOpen
  \bibfield  {author} {\bibinfo {author} {\bibfnamefont {M.}~\bibnamefont
  {Santha}}\ and\ \bibinfo {author} {\bibfnamefont {U.~V.}\ \bibnamefont
  {Vazirani}},\ }\href {\doibase
  http://dx.doi.org/10.1016/0022-0000(86)90044-9} {\bibfield  {journal}
  {\bibinfo  {journal} {Journal of Computer and System Sciences}\ }\textbf
  {\bibinfo {volume} {33}},\ \bibinfo {pages} {75 } (\bibinfo {year}
  {1986})}\BibitemShut {NoStop}%
\bibitem [{\citenamefont {{Colbeck}}\ and\ \citenamefont
  {{Renner}}(2012)}]{ColbeckRenner-Freerandomnesscan-2012}%
  \BibitemOpen
  \bibfield  {author} {\bibinfo {author} {\bibfnamefont {R.}~\bibnamefont
  {{Colbeck}}}\ and\ \bibinfo {author} {\bibfnamefont {R.}~\bibnamefont
  {{Renner}}},\ }\href {\doibase 10.1038/nphys2300} {\bibfield  {journal}
  {\bibinfo  {journal} {Nature Physics}\ }\textbf {\bibinfo {volume} {8}},\
  \bibinfo {pages} {450} (\bibinfo {year} {2012})},\ \Eprint
  {http://arxiv.org/abs/1105.3195} {arXiv:1105.3195 [quant-ph]} \BibitemShut
  {NoStop}%
\bibitem [{\citenamefont {{Gallego}}\ \emph {et~al.}(2013)\citenamefont
  {{Gallego}}, \citenamefont {{Masanes}}, \citenamefont {{de la Torre}},
  \citenamefont {{Dhara}}, \citenamefont {{Aolita}},\ and\ \citenamefont
  {{Ac{\'{\i}}n}}}]{GallegoMasanesEtAl-Fullrandomnessfrom-2013}%
  \BibitemOpen
  \bibfield  {author} {\bibinfo {author} {\bibfnamefont {R.}~\bibnamefont
  {{Gallego}}}, \bibinfo {author} {\bibfnamefont {L.}~\bibnamefont
  {{Masanes}}}, \bibinfo {author} {\bibfnamefont {G.}~\bibnamefont {{de la
  Torre}}}, \bibinfo {author} {\bibfnamefont {C.}~\bibnamefont {{Dhara}}},
  \bibinfo {author} {\bibfnamefont {L.}~\bibnamefont {{Aolita}}}, \ and\
  \bibinfo {author} {\bibfnamefont {A.}~\bibnamefont {{Ac{\'{\i}}n}}},\ }\href
  {\doibase 10.1038/ncomms3654} {\bibfield  {journal} {\bibinfo  {journal}
  {Nature Communications}\ }\textbf {\bibinfo {volume} {4}},\ \bibinfo {eid}
  {2654} (\bibinfo {year} {2013}),\ 10.1038/ncomms3654},\ \Eprint
  {http://arxiv.org/abs/1210.6514} {arXiv:1210.6514 [quant-ph]} \BibitemShut
  {NoStop}%
\bibitem [{\citenamefont {{Mironowicz}}\ and\ \citenamefont
  {{Pawlowski}}(2013)}]{MironowiczPawlowski-Amplificationofarbitrarily-2013}%
  \BibitemOpen
  \bibfield  {author} {\bibinfo {author} {\bibfnamefont {P.}~\bibnamefont
  {{Mironowicz}}}\ and\ \bibinfo {author} {\bibfnamefont {M.}~\bibnamefont
  {{Pawlowski}}},\ }\href@noop {} {\enquote {\bibinfo {title} {{Amplification
  of arbitrarily weak randomness}},}\ } (\bibinfo {year} {2013}),\ \Eprint
  {http://arxiv.org/abs/1301.7722} {arXiv:1301.7722 [quant-ph]} \BibitemShut
  {NoStop}%
\bibitem [{\citenamefont {{Grudka}}\ \emph {et~al.}(2013)\citenamefont
  {{Grudka}}, \citenamefont {{Horodecki}}, \citenamefont {{Horodecki}},
  \citenamefont {{Horodecki}}, \citenamefont {{Paw{\l}owski}},\ and\
  \citenamefont
  {{Ramanathan}}}]{GrudkaHorodeckiHorodeckiEtAl-Freerandomnessamplification-20%
13}%
  \BibitemOpen
  \bibfield  {author} {\bibinfo {author} {\bibfnamefont {A.}~\bibnamefont
  {{Grudka}}}, \bibinfo {author} {\bibfnamefont {K.}~\bibnamefont
  {{Horodecki}}}, \bibinfo {author} {\bibfnamefont {M.}~\bibnamefont
  {{Horodecki}}}, \bibinfo {author} {\bibfnamefont {P.}~\bibnamefont
  {{Horodecki}}}, \bibinfo {author} {\bibfnamefont {M.}~\bibnamefont
  {{Paw{\l}owski}}}, \ and\ \bibinfo {author} {\bibfnamefont {R.}~\bibnamefont
  {{Ramanathan}}},\ }\href@noop {} {\enquote {\bibinfo {title} {{Free
  randomness amplification using bipartite chain correlations}},}\ } (\bibinfo
  {year} {2013}),\ \Eprint {http://arxiv.org/abs/1303.5591} {arXiv:1303.5591
  [quant-ph]} \BibitemShut {NoStop}%
\bibitem [{\citenamefont {{Ramanathan}}\ \emph {et~al.}(2013)\citenamefont
  {{Ramanathan}}, \citenamefont {{Brandao}}, \citenamefont {{Grudka}},
  \citenamefont {{Horodecki}}, \citenamefont {{Horodecki}},\ and\ \citenamefont
  {{Horodecki}}}]{RamanathanBrandaoGrudkaEtAl-RobustDeviceIndependent-2013}%
  \BibitemOpen
  \bibfield  {author} {\bibinfo {author} {\bibfnamefont {R.}~\bibnamefont
  {{Ramanathan}}}, \bibinfo {author} {\bibfnamefont {F.~G.~S.~L.}\ \bibnamefont
  {{Brandao}}}, \bibinfo {author} {\bibfnamefont {A.}~\bibnamefont {{Grudka}}},
  \bibinfo {author} {\bibfnamefont {K.}~\bibnamefont {{Horodecki}}}, \bibinfo
  {author} {\bibfnamefont {M.}~\bibnamefont {{Horodecki}}}, \ and\ \bibinfo
  {author} {\bibfnamefont {P.}~\bibnamefont {{Horodecki}}},\ }\href@noop {}
  {\enquote {\bibinfo {title} {{Robust Device Independent Randomness
  Amplification}},}\ } (\bibinfo {year} {2013}),\ \Eprint
  {http://arxiv.org/abs/1308.4635} {arXiv:1308.4635 [quant-ph]} \BibitemShut
  {NoStop}%
\bibitem [{\citenamefont {{Brand{\~a}o}}\ \emph {et~al.}(2013)\citenamefont
  {{Brand{\~a}o}}, \citenamefont {{Ramanathan}}, \citenamefont {{Grudka}},
  \citenamefont {{Horodecki}}, \citenamefont {{Horodecki}},\ and\ \citenamefont
  {{Horodecki}}}]{BrandaoRamanathanGrudkaEtAl-RobustDevice-IndependentRandomne%
ss-2013}%
  \BibitemOpen
  \bibfield  {author} {\bibinfo {author} {\bibfnamefont {F.~G.~S.~L.}\
  \bibnamefont {{Brand{\~a}o}}}, \bibinfo {author} {\bibfnamefont
  {R.}~\bibnamefont {{Ramanathan}}}, \bibinfo {author} {\bibfnamefont
  {A.}~\bibnamefont {{Grudka}}}, \bibinfo {author} {\bibfnamefont
  {K.}~\bibnamefont {{Horodecki}}}, \bibinfo {author} {\bibfnamefont
  {M.}~\bibnamefont {{Horodecki}}}, \ and\ \bibinfo {author} {\bibfnamefont
  {P.}~\bibnamefont {{Horodecki}}},\ }\href@noop {} {\enquote {\bibinfo {title}
  {{Robust Device-Independent Randomness Amplification with Few Devices}},}\ }
  (\bibinfo {year} {2013}),\ \Eprint {http://arxiv.org/abs/1310.4544}
  {arXiv:1310.4544 [quant-ph]} \BibitemShut {NoStop}%
\bibitem [{\citenamefont {{Coudron}}\ and\ \citenamefont
  {{Yuen}}(2013)}]{CoudronYuen-InfiniteRandomnessExpansion-2013}%
  \BibitemOpen
  \bibfield  {author} {\bibinfo {author} {\bibfnamefont {M.}~\bibnamefont
  {{Coudron}}}\ and\ \bibinfo {author} {\bibfnamefont {H.}~\bibnamefont
  {{Yuen}}},\ }\href@noop {} {\enquote {\bibinfo {title} {{Infinite Randomness
  Expansion and Amplification with a Constant Number of Devices}},}\ }
  (\bibinfo {year} {2013}),\ \Eprint {http://arxiv.org/abs/1310.6755}
  {arXiv:1310.6755 [quant-ph]} \BibitemShut {NoStop}%
\bibitem [{\citenamefont {Koh}\ \emph {et~al.}(2012)\citenamefont {Koh},
  \citenamefont {Hall}, \citenamefont {Setiawan}, \citenamefont {Pope},
  \citenamefont {Marletto}, \citenamefont {Kay}, \citenamefont {Scarani},\ and\
  \citenamefont {Ekert}}]{KohHallSetiawanEtAl-EffectsofReduced-2012}%
  \BibitemOpen
  \bibfield  {author} {\bibinfo {author} {\bibfnamefont {D.~E.}\ \bibnamefont
  {Koh}}, \bibinfo {author} {\bibfnamefont {M.~J.~W.}\ \bibnamefont {Hall}},
  \bibinfo {author} {\bibnamefont {Setiawan}}, \bibinfo {author} {\bibfnamefont
  {J.~E.}\ \bibnamefont {Pope}}, \bibinfo {author} {\bibfnamefont
  {C.}~\bibnamefont {Marletto}}, \bibinfo {author} {\bibfnamefont
  {A.}~\bibnamefont {Kay}}, \bibinfo {author} {\bibfnamefont {V.}~\bibnamefont
  {Scarani}}, \ and\ \bibinfo {author} {\bibfnamefont {A.}~\bibnamefont
  {Ekert}},\ }\href {\doibase 10.1103/PhysRevLett.109.160404} {\bibfield
  {journal} {\bibinfo  {journal} {Phys. Rev. Lett.}\ }\textbf {\bibinfo
  {volume} {109}},\ \bibinfo {pages} {160404} (\bibinfo {year}
  {2012})}\BibitemShut {NoStop}%
\bibitem [{\citenamefont {Thinh}\ \emph {et~al.}(2013)\citenamefont {Thinh},
  \citenamefont {Sheridan},\ and\ \citenamefont
  {Scarani}}]{ThinhSheridanScarani-Belltestswith-2013}%
  \BibitemOpen
  \bibfield  {author} {\bibinfo {author} {\bibfnamefont {L.~P.}\ \bibnamefont
  {Thinh}}, \bibinfo {author} {\bibfnamefont {L.}~\bibnamefont {Sheridan}}, \
  and\ \bibinfo {author} {\bibfnamefont {V.}~\bibnamefont {Scarani}},\ }\href
  {\doibase 10.1103/PhysRevA.87.062121} {\bibfield  {journal} {\bibinfo
  {journal} {Phys. Rev. A}\ }\textbf {\bibinfo {volume} {87}},\ \bibinfo
  {pages} {062121} (\bibinfo {year} {2013})}\BibitemShut {NoStop}%
\bibitem [{\citenamefont {{Plesch}}\ and\ \citenamefont
  {{Pivoluska}}(2013)}]{PleschPivoluska-SingleMin-EntropyRandom-2013}%
  \BibitemOpen
  \bibfield  {author} {\bibinfo {author} {\bibfnamefont {M.}~\bibnamefont
  {{Plesch}}}\ and\ \bibinfo {author} {\bibfnamefont {M.}~\bibnamefont
  {{Pivoluska}}},\ }\href@noop {} {\enquote {\bibinfo {title} {{Single
  Min-Entropy Random Sources can be Amplified}},}\ } (\bibinfo {year} {2013}),\
  \Eprint {http://arxiv.org/abs/1305.0990} {arXiv:1305.0990 [quant-ph]}
  \BibitemShut {NoStop}%
\bibitem [{\citenamefont {{Chung}}\ \emph {et~al.}(2014)\citenamefont
  {{Chung}}, \citenamefont {{Shi}},\ and\ \citenamefont
  {{Wu}}}]{ChungShiWu-PhysicalRandomnessExtractors-2014}%
  \BibitemOpen
  \bibfield  {author} {\bibinfo {author} {\bibfnamefont {K.-M.}\ \bibnamefont
  {{Chung}}}, \bibinfo {author} {\bibfnamefont {Y.}~\bibnamefont {{Shi}}}, \
  and\ \bibinfo {author} {\bibfnamefont {X.}~\bibnamefont {{Wu}}},\ }\href@noop
  {} {\enquote {\bibinfo {title} {{Physical Randomness Extractors}},}\ }
  (\bibinfo {year} {2014}),\ \Eprint {http://arxiv.org/abs/1402.4797}
  {arXiv:1402.4797 [quant-ph]} \BibitemShut {NoStop}%
\bibitem [{\citenamefont {Mermin}(1990)}]{Mermin1990}%
  \BibitemOpen
  \bibfield  {author} {\bibinfo {author} {\bibfnamefont {N.~D.}\ \bibnamefont
  {Mermin}},\ }\href {\doibase 10.1103/PhysRevLett.65.1838} {\bibfield
  {journal} {\bibinfo  {journal} {Phys. Rev. Lett.}\ }\textbf {\bibinfo
  {volume} {65}},\ \bibinfo {pages} {1838} (\bibinfo {year}
  {1990})}\BibitemShut {NoStop}%
\bibitem [{\citenamefont {{Navascu{\'e}s}}\ \emph {et~al.}(2008)\citenamefont
  {{Navascu{\'e}s}}, \citenamefont {{Pironio}},\ and\ \citenamefont
  {{Ac{\'{\i}}n}}}]{Navascues2008}%
  \BibitemOpen
  \bibfield  {author} {\bibinfo {author} {\bibfnamefont {M.}~\bibnamefont
  {{Navascu{\'e}s}}}, \bibinfo {author} {\bibfnamefont {S.}~\bibnamefont
  {{Pironio}}}, \ and\ \bibinfo {author} {\bibfnamefont {A.}~\bibnamefont
  {{Ac{\'{\i}}n}}},\ }\href {\doibase 10.1088/1367-2630/10/7/073013} {\bibfield
   {journal} {\bibinfo  {journal} {New Journal of Physics}\ }\textbf {\bibinfo
  {volume} {10}},\ \bibinfo {eid} {073013} (\bibinfo {year} {2008})},\ \Eprint
  {http://arxiv.org/abs/0803.4290} {arXiv:0803.4290 [quant-ph]} \BibitemShut
  {NoStop}%
\bibitem [{\citenamefont {Scarani}\ \emph {et~al.}(2006)\citenamefont
  {Scarani}, \citenamefont {Gisin}, \citenamefont {Brunner}, \citenamefont
  {Masanes}, \citenamefont {Pino},\ and\ \citenamefont
  {Ac\'in}}]{ScaraniGisinBrunnerEtAl-Secrecyextractionfrom-2006}%
  \BibitemOpen
  \bibfield  {author} {\bibinfo {author} {\bibfnamefont {V.}~\bibnamefont
  {Scarani}}, \bibinfo {author} {\bibfnamefont {N.}~\bibnamefont {Gisin}},
  \bibinfo {author} {\bibfnamefont {N.}~\bibnamefont {Brunner}}, \bibinfo
  {author} {\bibfnamefont {L.}~\bibnamefont {Masanes}}, \bibinfo {author}
  {\bibfnamefont {S.}~\bibnamefont {Pino}}, \ and\ \bibinfo {author}
  {\bibfnamefont {A.}~\bibnamefont {Ac\'in}},\ }\href {\doibase
  10.1103/PhysRevA.74.042339} {\bibfield  {journal} {\bibinfo  {journal} {Phys.
  Rev. A}\ }\textbf {\bibinfo {volume} {74}},\ \bibinfo {pages} {042339}
  (\bibinfo {year} {2006})}\BibitemShut {NoStop}%
\bibitem [{\citenamefont {Naor}\ and\ \citenamefont
  {Naor}(1993)}]{NaorNaor-Small-BiasProbabilitySpaces:-1993}%
  \BibitemOpen
  \bibfield  {author} {\bibinfo {author} {\bibfnamefont {J.}~\bibnamefont
  {Naor}}\ and\ \bibinfo {author} {\bibfnamefont {M.}~\bibnamefont {Naor}},\
  }\href
  {http://dblp.uni-trier.de/db/journals/siamcomp/siamcomp22.html#NaorN93}
  {\bibfield  {journal} {\bibinfo  {journal} {SIAM J. Comput.}\ }\textbf
  {\bibinfo {volume} {22}},\ \bibinfo {pages} {838} (\bibinfo {year}
  {1993})}\BibitemShut {NoStop}%
\end{thebibliography}%

\appendix

\section{Construction of the class $H$}

\label{sec:construction_of_hash_functions}
Let $H=\{h_{1},\dots,h_{m}\}$ be a family of hash functions s.t.
$h_{i}:\{0,...,N-1\}\rightarrow\{0,1,2,3\}$. Let us assume that we receive an
element of $\{0,\dots,N-1\}$ drawn randomly according some (non-uniform)
distribution with min-entropy $\log_{2}k$ (we consider only $k\geq4$).

We want to construct $H$ with the property that for every set $S\subseteq
\{0,\dots,N-1\}$ with $\left\vert S\right\vert \geq k$ there is at least one
hash function $h\in H$ such that $h(S)=\{0,1,2,3\}$. This is trivially
satisfied for $H_{full}=\{0,1,2,3\}^{N}$, however, such a class of functions
is unpractically large, it has $4^{N}$ elements. Therefor we shall construct a
smaller set fulfilling the condition.


\subsection{Derandomization construction of the class $H$}

Let us consider a sequence of random variables $Z=\left(  Z_{0},\dots
,Z_{N-1}\right)  $ such that $Z_{i}\in\{0,1,2,3\}$. The outcomes of such a
random experiment are $N$-position sequences from the set $\{0,1,2,3\}^{N}$.
It is easy to see that each such sequence specifies uniquely a particular
function $h:\{0,...,N-1\}\rightarrow\{0,1,2,3\}$, and vice versa. Since now on
we will use them interchangeably.

Let us assume that random variables $Z$ satisfy the condition that for every
$4$--tuple of positions $j_{0},j_{1},j_{2},j_{3}$ and every $4$-element string
$a_{0}a_{1}a_{2}a_{3}\in\{0,1,2,3\}^{4}$ it holds that
\begin{equation}
P\bigl[Z_{j_{0}}=a_{0}\wedge Z_{j_{1}}=a_{1}\wedge Z_{j_{2}}=a_{2}\wedge
Z_{j_{3}}=a_{3}\bigr]>0. \label{Eq1}%
\end{equation}
Note that for our purposes even a weaker assumption on $Z$ is sufficient: It
is enough if for every $4$--tuple of positions $j_{0},j_{1},j_{2},j_{3}$ there
exists at least one $4$-element string $a_{0}a_{1}a_{2}a_{3}\in\{0,1,2,3\}^{4}%
$ with all $a_{0},a_{1},a_{2},a_{3}$ begin mutually different and satisfying
(\ref{Eq1}). However, the stronger condition will make it easier to find a
suitable set.

Let us denote $H=\{a\in\{0,1,2,3\}^{N}\text{ s.t. }P[Z=a]>0\}$. Using the
probabilistic method we see, that for each $4$--tuple of positions
$j_{0},j_{1},j_{2},j_{3}$ and every $4$-element string $a_{0}a_{1}a_{2}%
a_{3}\in\{0,1,2,3\}^{4}$ there exists a function $h\in H$ such that
\[
h(j_{0})=a_{0}\wedge h(j_{1})=a_{1}\wedge h(j_{2})=a_{2}\wedge h(j_{3}%
)=a_{3}.
\]
The number of functions in $H$ is the same as the number of (nonzero
probability) sample space elements of $Z$. It remains to construct $Z$ with a
sample space as small as possible.


\subsection{Construction of a random variable $Z$}


\begin{definition}
We define the distance of two distributions $D_{1}$ and $D_{2}$ by
\[
||D_{1}-D_{2}||=\sum_{\omega\in\Omega}\left\vert D_{1}(\omega)-D_{2}%
(\omega)\right\vert ,
\]
where $\Omega$ is the set of all possible events.
\end{definition}

\begin{definition}
Binary random variables are \textbf{$k$-wise $\delta$-dependent} iff for all
subsets $S\subseteq\{0,\dots,N-1\},\left\vert S\right\vert \leq k$
\[
||U(S)-D(S)||\leq\delta,
\]

\end{definition}
where $U(S)$ is a uniform distribution over $|S|$-bit strings and $D(S)$ is a
marginal distribution over subset of variables specified by $S$.

\begin{theorem}
\label{thm} The logarithm of the cardinality of the sample space needed for
constructing $N$ $k$-wise $\delta$-dependent random variables is $O\left(
k+\log\log N+\log\frac{1}{\delta}\right) $
\cite{NaorNaor-Small-BiasProbabilitySpaces:-1993}.
\end{theorem}

Let us consider two sequences $X_{0},\dots,X_{N-1}$ and $Y_{0},\dots,Y_{N-1}$
of binary $4$-wise $\delta$--dependent random variables, both sequences being
mutually independent. Let $Z_{i}=2X_{i}+Y_{i}$.

As both $X$ and $Y$ are $\delta$-dependent, their distance from the uniform
distribution on every subset of size at most $4$ is at most $\delta$. Assuming
there is a zero probability for at least one binary string out of
$\{0,1\}^{4}$ at positions $(0,1,2,3)$ we have that the distance of such a
distribution from the uniform distribution is at least $2\times2^{-4}=2^{-3}%
$.

Hence, assuring that $\delta<2^{-3}$ we obtain that for each $4$ positions
there is a nonzero probability of every $4$-bit sequence appearing. Hence, for
the sequence of random variables $Z$ it holds that every $4$-tuple of
positions every string out of $\{0,1,2,3\}^{4}$ appears with non-zero probability.

In our case we need two independent sets of $N=2^{n}$ $4$-wise $1/8$-dependent
random variables, resulting in a sample space of $O(n^{c})$, bearing the
desired polynomial construction.

\section{Robustness}

Let us assume we would tolerate a failure at most $\frac{\left(
1-f(\varepsilon)\right)  }{2}l$ devices during the run of the whole protocol.
Let us first calculate the number of rounds of the protocol $l$ needed to
obtain the original $\varepsilon$ and $\delta$ characteristics of the
non-robust device.

\subsection{Efficiency}

Assuming the adversary is cheating (wants to achieve bias greater than
$\varepsilon$), in each round of the protocol there will be at least one
device failure with probability $1-f(\varepsilon)$. The probability $\delta$
that the adversary stays undetected while all devices produce bias at least
$\varepsilon$ is bounded by the distribution function of the binomial
distribution
\[
\delta\leq F\left(  \frac{\left(  1-f(\varepsilon)\right)  }{2}%
l;l;1-f(\varepsilon)\right)  .
\]
This probability can be upper bounded by Chernoff's inequality by
\begin{equation}
F\left(  \frac{\left(  1-f(\varepsilon)\right)  }{2}l;l;1-f(\varepsilon
)\right)  \leq e^{-\frac{\left(  1-f(\varepsilon)\right)  }{8}l}.
\end{equation}
We can derive the necessary number of rounds of the protocol $l$ to be
\[
l>8\frac{\ln\delta}{f(\varepsilon)-1}.
\]
Comparing to the number of rounds needed for the non-robust protocol
$\frac{\log\delta}{\log f(\varepsilon)}$ we can obtain the scaling factor $s$
to be%
\[
s=8\frac{\ln f(\varepsilon)}{f(\varepsilon)-1}.
\]
For $f(\varepsilon)\rightarrow1$ (what is the case for small $\varepsilon$)
the scaling factor approaches a constant of $8$.

\subsection{Imperfectness}

We also want to assure there exist a non-zero failing probability of each
individual device $\mu$ such that the protocol execution will not be (falsely)
declared to be attacked by the adversary with high probability. Let us
consider an honest provider (not trying to cheat) and set $\mu=\frac
{1-f(\varepsilon)}{4m}$. We will calculate the probability that more than
$\frac{\left(  1-f(\varepsilon)\right)  }{2}l$ devices will fail during the process.

Since the producer of the devices is assumed to be honest (otherwise the
protocol failure is justified), we may assume that failures of devices are
independent of each other. Therefore the failures can be modeled by i.i.d.
Bernoulli random variables ($Z_{i}=1$ if the $i$--th device fails the test)
$Z_{1},\dots,Z_{ml}$, with $P(Z_{i}=1)=\mu=\frac{1-f(\varepsilon)}{4m})$. The
number of failures $Z=\sum_{i=1}^{ml}Z_{i}$ is binomially distributed. For the
protocol not to abort we need less than $\frac{1-f(\epsilon)}{2}l$ failures,
hence we need to upper bound the probability
\[
P\left(  \sum_{i=1}^{ml}Z_{i}>\frac{1-f(\epsilon)}{2}l\right)  =F\left(
ml-l\frac{1-f(\epsilon)}{2},ml,1-\mu\right)  .
\]

We can use the Hoeffding inequality:%

\[
P\left(  \sum_{i=1}^{ml}Z_{i}>\frac{1-f(\epsilon)}{2}l\right)  \leq
e^{-\frac{\left(  1-f(\varepsilon)\right)  ^{2}}{8m}l},
\]
i.e. the probability of false protocol abort drops exponentially with the
number of rounds $l$.


\section{Using $H_{full}$ for Non-Block Sources}

We used the following claim in the main text:
If we hash the outcome of a $(n,Rn)$-flat distribution by each of the hash functions
from the full set of functions $H_{full} = \left\{h_i: \{0,1\}^n\mapsto\{0,1,2,3\}\right\}$,
at least $\frac{Rn}{2}$ functions have uniform and independent outcomes.

First let us suppose $Rn$ is natural and even. Then there are $4^{Rn/2}$ strings appearing
with probability $\frac{1}{4^{Rn/2}}$. Let us label them
$\{s_i\}_{i = 0}^{(4^{Rn/2}-1)}$. We will now explicitly construct hash functions
$\{h_j\}_{j=0}^{Rn/2}$ with desired properties.

Let $M$ be $\frac{Rn}{2}$ times $4^{Rn/2}$ matrix with $i^{th}$ column
being a representation of $i$ in base $4$. Let us assign $h_j(s_i) = M_{ji}$
(example with $Rn = 4$ is depicted in Fig. (\ref{matrix})).
Although this is only a partial definition of $\{h_j\}_{j=0}^{Rn/2}$, it is sufficient
for our purposes, because other strings appear with probability 0.
It should now be straightforward to see that each vector of outcomes
$(h_0,\dots,h_{Rn/2})$ appears with equal probability and therefore marginal
distributions of outputs of a single function $h$ is uniform and independent on the others.

\begin{figure}[h!]
\begin{tabular}{c|cccccccccccccccc}
& $s_0$ & $s_1$ & $s_2$ & $s_3$ & $s_4$ & $s_5$ & $s_6$ & $s_7$
& $s_8$ & $s_8$ & $s_{10}$ & $s_{11}$ & $s_{12}$  & $s_{13}$ & $s_{14}$ & $s_{15}$ \\\hline
$h_0$ & 0 & 0 & 0 & 0 & 1 & 1 & 1 & 1 & 2 & 2 & 2 & 2 & 3 & 3 & 3 & 3\\
$h_1$ & 0 & 1 & 2 & 3 & 0 & 1 & 2 & 3 & 0 & 1 & 2 & 3 & 0 & 1 & 2 & 3\\
\end{tabular}
\caption{Matrix $M$ for $Rn = 4$.}
\label{matrix}
\end{figure}

By Caratheodory theorem all other values of $Rn$ can be written as convex combinations
of $(n,m)$ flat sources with $m = 2\lfloor Rn/2 \rfloor$, which gives us that the probability
to cheat with such $(n,Rn)$ source is at most the same as with $(n,m)$ flat source --
i.~e. equal to $\lfloor \frac{Rn}{2}\rfloor$ boxes obtaining uniform independent inputs.

\end{document}